\documentclass[11pt]{article}
\usepackage[a4paper,margin=1in]{geometry}
\usepackage{amsmath,amssymb,booktabs,graphicx,microtype}
\usepackage[numbers,sort&compress]{natbib}
\usepackage[colorlinks=true,allcolors=blue!55!black]{hyperref}
\usepackage{xcolor}
\usepackage{enumitem,float}
\usepackage{placeins}
\usepackage{algorithm}
\usepackage{algpseudocode}

\newcommand{\ResidentGaussians}{2.7}
\newcommand{\ResidentVRAM}{72.26}
\newcommand{\ResidentUtilMean}{99.9}
\newcommand{\ResidentUtilPctl}{100.0}
\newcommand{\ResidentPower}{288.3}
\newcommand{\ResidentEnergyKJ}{10.82}
\newcommand{\SweepThroughput}{360.46}
\newcommand{\SweepUtilization}{98.21}
\newcommand{\SweepMemoryGiB}{67.30}
\newcommand{\SweepPower}{279.5}
\newcommand{\RooflineLargeSpeedupThirtyTwo}{3.99}

\newcommand{\RooflineLargeBandwidthPercent}{8.45}

\newcommand{\UtilityStates}{32768}
\newcommand{\UtilitySeconds}{17.34}
\newcommand{\UtilityThroughput}{1889.7}
\newcommand{\UtilityVRAM}{21.07}

\newcommand{\RealUtilityClusters}{32}
\newcommand{\RealUtilityContexts}{32}

\newcommand{\RealUtilityFineNegativeMean}{3}
\newcommand{\RealUtilityFineNegativeTail}{21}
\newcommand{\RealUtilityFineCINegative}{2}
\newcommand{\BroomUtilityClusters}{64}
\newcommand{\BroomUtilityContexts}{16}

\newcommand{\BroomUtilityFineNegativeMean}{49}
\newcommand{\BroomUtilityFineNegativeTail}{61}
\newcommand{\BroomUtilityFineCINegative}{43}
\newcommand{\BroomTrainFineNegativeMean}{51}
\newcommand{\BroomTrainFineNegativeTail}{61}
\newcommand{\ChickenUtilityClusters}{64}
\newcommand{\ChickenUtilityContexts}{64}

\newcommand{\ChickenUtilityFineNegativeMean}{28}
\newcommand{\ChickenUtilityFineNegativeTail}{42}
\newcommand{\ChickenUtilityFineCINegative}{22}
\newcommand{\ChickenSourcePSNR}{28.03}
\newcommand{\ChickenPrefixZeroPSNR}{21.41}
\newcommand{\ChickenPrefixOnePSNR}{25.17}
\newcommand{\ChickenPrefixTwoPSNR}{27.48}
\newcommand{\ChickenPrefixThreePSNR}{27.70}
\newcommand{\ChickenPrefixViolations}{0}
\newcommand{\ChickenTrainedCleanGain}{0.11}

\newcommand{\ChickenPairedPSNRGain}{0.078}
\newcommand{\ChickenPairedPSNRCILow}{0.075}
\newcommand{\ChickenPairedPSNRCIHigh}{0.080}
\newcommand{\ChickenPairedWinPercent}{100}
\newcommand{\ChickenPairedLPIPSGain}{0.33}
\newcommand{\ChickenPairedTemporalLPIPSGain}{0.35}
\newcommand{\ChickenPairedFlickerGain}{1.00}
\newcommand{\ChickenTrainedPrefixViolations}{0}
\newcommand{\ChickenCrossviewPSNRGain}{3.03}
\newcommand{\ChickenCrossviewLPIPSReduction}{12.88}
\newcommand{\ChickenCrossviewLayerBudgetUse}{99.1}
\newcommand{\ChickenRobustCrossviewPSNRGain}{3.05}
\newcommand{\ChickenRobustCrossviewLPIPSReduction}{12.80}

\newcommand{\ChickenBurstPSNRGain}{0.064}
\newcommand{\ChickenBurstPSNRCILow}{0.061}
\newcommand{\ChickenBurstPSNRCIHigh}{0.067}
\newcommand{\ChickenBurstLPIPSGain}{0.30}
\newcommand{\ChickenBurstTemporalLPIPSGain}{0.32}
\newcommand{\ChickenBurstFlickerGain}{0.82}
\newcommand{\ChickenBurstReferenceCliff}{14.06}
\newcommand{\ChickenBurstConditionedCliff}{14.84}
\newcommand{\FrontierUtilityViolations}{0}
\newcommand{\FrontierLayerViolations}{2}
\newcommand{\FrontierLowPSNRGain}{0.33}
\newcommand{\FrontierLowLPIPSReduction}{1.77}
\newcommand{\FrontierFinalPSNRGain}{0.07}
\newcommand{\FrontierFinalLPIPSReduction}{0.79}
\newcommand{\FrontierLayerBudgetUse}{98.7}
\newcommand{\RobustFrontierUtilityViolations}{0}
\newcommand{\RobustFrontierLayerViolations}{1}
\newcommand{\RobustFrontierFinalPSNRGain}{0.12}
\newcommand{\RobustFrontierFinalLPIPSReduction}{0.64}

\newcommand{\ErasureCleanPSNRCost}{0.15}
\newcommand{\ErasureTenDropUntrained}{0.72}
\newcommand{\ErasureTenDropTrained}{0.65}
\newcommand{\ErasureTenAbsolutePSNRDeficit}{0.08}
\newcommand{\ErasureTenLPIPSDeficit}{0.21}
\newcommand{\ErasureBurstPSNRGain}{0.09}
\newcommand{\ErasureBurstLPIPSGain}{0.25}
\newcommand{\TrainingGaussians}{0.25}
\newcommand{\TrainingWorlds}{96}
\newcommand{\TrainingSteps}{240}

\newcommand{\TrainingDeadlineCliff}{100}
\newcommand{\SelectedTemporalRank}{2}
\newcommand{\RankCleanGainOverFour}{0.18}
\newcommand{\RankBurstGainOverFour}{0.29}
\newcommand{\TransportSimulations}{44800}
\newcommand{\TransportTrajectoryRows}{6764800}
\newcommand{\TransportTraceCount}{7}
\newcommand{\TransportRealPairs}{600}
\newcommand{\TransportMissReduction}{25.6}
\newcommand{\TransportMissReductionLow}{23.6}
\newcommand{\TransportMissReductionHigh}{27.4}
\newcommand{\TransportQualityGain}{2.2}
\newcommand{\TransportWasteReduction}{29.6}
\newcommand{\TransportStartupCost}{4.1}
\newcommand{\TransportStartupFailures}{1}
\newcommand{\AtlasBBUtilityLow}{32.49}
\newcommand{\AtlasBBNominalLow}{28.27}

\newcommand{\AtlasBBOracleLow}{30.27}

\newcommand{\AtlasJJUtilityLow}{29.27}
\newcommand{\AtlasJJNominalLow}{27.20}

\newcommand{\AtlasJJOracleLow}{27.25}

\newcommand{\AtlasBroomUtilityTop}{18.10}

\newcommand{\AtlasBroomOracleGapTop}{1.13}
\newcommand{\AtlasStatesEvaluated}{8,197}
\newcommand{\AtlasBudgetCount}{four}

\newcommand{\method}{the proposed packet-conditioned codec}
\newcommand{\loss}{\mathcal{L}}

\title{\textbf{Renderable Partial Representations for Dynamic Gaussian Splatting\\under Incomplete Delivery}}
\author{Faruk Alpay$^{1*}$ \quad Levent Sar{\i}o\u{g}lu$^{1}$ \quad Yaser Hadri$^{2}$\\
$^{1}$Department of Computer Engineering \quad $^{2}$Department of Artificial Intelligence\\
Bah\c{c}e\c{s}ehir University, Istanbul, Turkey\\
\texttt{\char`\{faruk.alpay, levent.sarioglu\char`\}@bahcesehir.edu.tr}\\
\texttt{yaser.hadri@bahcesehir.edu.tr}\\[0.35em]
\small $^{*}$Correspondence: \texttt{alpay@lightcap.ai}}
\date{}

\begin{document}
\maketitle

\begin{abstract}
Dynamic Gaussian compression is normally optimized for complete files or complete
progressive prefixes. Interactive rendering instead encounters partial
representations: some spatiotemporal regions are present, others are missing, and
late refinements cannot affect the displayed frame. We study dynamic Gaussian
representations whose incomplete delivery states remain directly renderable and
whose degradation is optimized in image space. Gaussian primitives are organized
into independently addressable spatiotemporal clusters with a base level and three
refinements. Cluster motion combines affine prediction with a compact temporal
basis selected by validation and quantized residuals. Training samples partial dependency graphs, renders
many counterfactual states in one GPU batch, and minimizes expected distortion,
tail distortion, temporal inconsistency, rate, and prefix regressions. A
counterfactual utility layer measures the marginal render contribution of each
completion group across valid receiver contexts. The same graph admits a concrete
delivery realization with MTU-bounded entropy-coded chunks, deadline-aware
scheduling, selective recovery, and receiver-side dependency closure. The H100
renderer evaluates these counterfactual states in batches and is bit-equivalent to
serial evaluation. On held-out views, the finest refinement has negative mean
marginal utility in \RealUtilityFineNegativeMean{}/\RealUtilityClusters{} D-NeRF
\texttt{bouncingballs} clusters and
\BroomUtilityFineNegativeMean{}/\BroomUtilityClusters{} HyperNeRF
\texttt{broom2} clusters, and
\ChickenUtilityFineNegativeMean{}/\ChickenUtilityClusters{} HyperNeRF
\texttt{chicken} clusters; its lower-tail utility is negative in
\RealUtilityFineNegativeTail{}/\RealUtilityClusters{} and
\BroomUtilityFineNegativeTail{}/\BroomUtilityClusters{} and
\ChickenUtilityFineNegativeTail{}/\ChickenUtilityClusters{} clusters, respectively.
On \texttt{broom2}, render-utility ordering removes both PSNR regressions produced
by nominal layer order at matched byte budgets. On \texttt{chicken}, utilities
measured on disjoint training cameras improve held-out PSNR by
\ChickenCrossviewPSNRGain{} dB at the lowest matched budget. These scoped results show why nominal refinement order cannot
substitute for render-conditioned utility. The resulting formulation treats
network delivery as a distribution over renderable scene states rather than as an
external wrapper around a graphics codec.
\end{abstract}

\section{Introduction}
Dynamic Gaussian splatting enables interactive free-viewpoint rendering, but its
scene state remains large enough that incomplete delivery changes the scene that
is actually rendered.
Prior work compresses temporal deformation~\citep{wu2024four,queen2024,light4gs2025},
builds progressive dynamic representations~\citep{progressiveDynamic2026}, accelerates GPU
encoding and decoding~\citep{gsnfs2026}, and adapts layered streams to estimated
throughput~\citep{lts2025}. These results establish that dynamic Gaussian delivery
must be progressive and fast. They do not imply that a receiver obtains every
packet in a complete layer prefix.

A viewer may therefore see a valid but incomplete Gaussian state: a moving region
without its appearance refinement, a base cluster without fine motion residuals,
or a temporally inconsistent mixture of refinement levels. Conventional
rate--distortion training does not expose these states because it evaluates
complete files or complete prefixes. The resulting failure is graphical before it
is network-level: local detail disappears abruptly, motion residuals become
temporally unstable, and disoccluded regions can collapse when one dependency is
missing.

The distinction from existing progressive dynamic Gaussian representations is not
progressiveness itself. Our hypothesis is that the representation should be
optimized over the partial dependency states from which images will actually be
rendered. Delivery deadlines and erasures define a distribution over those states;
they are not merely labels attached after graphics optimization.
Accordingly, the contribution is not dynamic Gaussian streaming in isolation, but
render-conditioned utility optimization over the incomplete scene states induced
by delivery.

The implemented contributions are:
\begin{enumerate}[leftmargin=1.6em,itemsep=2pt]
\item counterfactual render utility, which measures a completion group's marginal
image-space contribution across valid partial receiver contexts and establishes
the Renderable-State Utility Principle: nominal refinement order does not determine
utility under incomplete delivery;
\item optimization over a delivery-induced distribution of renderable partial
scene states, including expected distortion, lower-tail risk, temporal stability,
rate, and prefix monotonicity; and
\item and, as supporting infrastructure rather than a standalone systems claim, the
representation, batched renderer, packet format, sender scheduler, and receiver
closure needed to test this principle end to end, from graphics quality through
trace-driven delivery QoE.
\end{enumerate}

\section{Related Work and Scope}
\paragraph{Dynamic Gaussian compression.}
Wu et al. represent deformation with a compact spatiotemporal field~\citep{wu2024four}.
QUEEN quantizes frame-to-frame Gaussian residuals for streaming free-viewpoint
video~\citep{queen2024}; Light4GS combines spatiotemporal pruning and learned
context coding~\citep{light4gs2025}. GS-NFS targets full-frame-rate GPU compression
and decompression~\citep{gsnfs2026}. These systems motivate our temporal
factorization and entropy-model ablations, but our primary target is graceful
degradation under incomplete delivery rather than full-stream throughput alone.

\paragraph{Progressive streaming.}
LapisGS introduced cumulative layers for static Gaussian scenes~\citep{lapis2025}.
LTS combines layering, tiling, and DASH adaptation for dynamic scenes~\citep{lts2025}.
Li et al. decompose a dynamic representation into renderable prefixes
\citep{progressiveDynamic2026}. \method{} does not claim to be the first progressive dynamic
Gaussian codec. It changes the optimization and scheduling unit from a complete
layer prefix to a deadline-constrained dependency-completion group.

\paragraph{Transport.}
QUIC provides reliable streams and shared congestion control~\citep{quic9000},
with standardized loss detection and an exemplary NewReno controller
\citep{quic9002}. QUIC DATAGRAM carries unreliable application data without
retransmission while still consuming the connection congestion budget
\citep{quic9221}. Our emulator follows these interfaces and accounting rules, but
is not presented as a full QUIC stack. Native protocol experiments remain a
separate evaluation layer.

\section{Renderable Spatiotemporal Representation}
Let a fixed-identity dynamic scene contain $N$ Gaussians over $T$ timestamps. We
cluster primitives using normalized canonical position, displacement and velocity
statistics, and visibility history. A deterministic variance split bounds cluster
size and gives stable cluster identifiers. Every dependency-closed subset decodes
to a valid Gaussian scene, even when spatial regions use different refinement
levels.

For Gaussian $i$ in cluster $c$, position is represented as
\begin{equation}
\mu_{i,t}=A_{c,t}\bar\mu_i+\sum_{r=1}^{R}b_{t,r}q_{i,r}
+\delta^{(2)}_{i,t}+\delta^{(3)}_{i,t},
\end{equation}
where $A_{c,t}$ is a cluster affine transform, $b$ is a shared temporal basis,
$q$ contains per-Gaussian coefficients, and $\delta$ contains two residual levels.
Scale, rotation, opacity, and colour use a base plus three learned refinements.
The base preserves coarse geometry, opacity, and view-independent appearance;
later levels add motion residuals, higher-order appearance, and finer quantization
planes. Quantization steps are learned with a straight-through estimator. This
factorization exposes the visual cost of omitted content during optimization
rather than treating missing bytes as a decoder error.

\begin{figure}[H]
\centering
\includegraphics[width=0.98\linewidth]{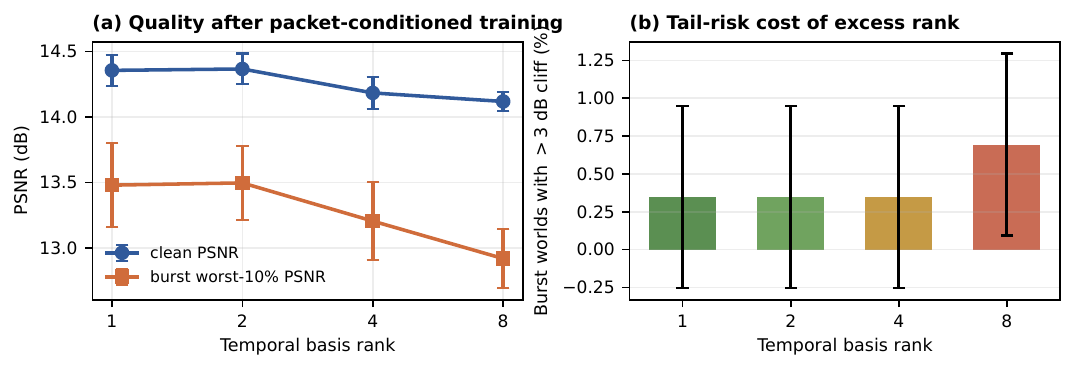}
\caption{Temporal-basis rank after identical packet-conditioned training on a
high-complexity controlled motion sequence (mean and standard deviation over
three seeds). Rank is a rate--quality--robustness choice, not a fixed architectural
constant.}
\label{fig:rank}
\end{figure}

We select $R$ from a small validation sweep rather than fixing rank four by
convention. On a four-mode nonrigid controlled sequence, identical
packet-conditioned training selects $R=\SelectedTemporalRank{}$: relative to
$R=4$, it improves clean PSNR by \RankCleanGainOverFour{} dB and burst
worst-10\% PSNR by \RankBurstGainOverFour{} dB. Figure~\ref{fig:rank} shows that
excess rank increases quantization and tail-risk cost without improving the
rendered sequence.

\FloatBarrier

Every cluster produces one base completion group and three refinement groups. Raw
signed 16-bit symbols are split into bounded chunks and encoded using a normalized
byte-wise rANS table. A DGSP media header identifies packet, cluster, layer,
temporal coverage, quantization profile, and dependencies. CRC32 rejects corrupted
chunks. Payload is limited to 1200 bytes so the transport can remain below common
Internet MTUs after transport overhead.

The human-readable JSON manifest is an artifact format, not an on-wire requirement.
Transport uses a compressed, CRC-protected DGSM control manifest on a reliable
stream. Media received before control completion may be buffered but is not marked
renderable. This separation makes manifest cost part of startup latency rather than
hiding it in offline setup.

\section{Packet-Conditioned Rendering Objective}
A training world samples capacity, RTT, display deadline, IID erasure, and a
Gilbert--Elliott burst process. Packet dependencies are enforced after delivery,
then converted to cluster-layer masks. After a 20\% reconstruction warm-up,
optimization minimizes
\begin{equation}
\loss=\mathbb{E}_{w}[D_w]+\lambda_c\operatorname{CVaR}_{0.05}(D)
+\lambda_rR+\lambda_tF+\lambda_mM,
\end{equation}
where $D$ combines RGB and SSIM loss, $R$ is a differentiable rate proxy, $F$
penalizes temporal-change error, and $M$ penalizes quality regressions as a prefix
grows. Complete prefixes receive direct supervision as an anchor, while partial
dependency states provide the robustness signal.

The reported objective-selection artifact fixes all weights rather than tuning
them after reading the final curves. It selects the zero-violation run with the
highest held-out partial-state p10 PSNR, breaking ties by lower wire bytes:
$D=\mathrm{MSE}+0.20(1-\mathrm{SSIM})$, $\lambda_c=0.30$,
$\lambda_r=0$, $\lambda_t=0.10$, and $\lambda_m=2.0$, with a separate complete
prefix anchor weight of 0.50. The rate proxy $R$ is the logical completion-group
wire cost before rANS: raw stream bytes plus one 48-byte chunk header and 4 bytes
per dependency word for each 1200-byte payload chunk. This makes equal-wire-rate
comparisons reproducible from the packet graph.

A three-seed sensitivity study evaluates CVaR tail fractions 0.05, 0.10, and
0.20 under the same masks and optimization budget. Their held-out partial-state
p10 PSNR values are 26.059, 26.057, and 26.057 dB, respectively, with zero
aggregate prefix violations for every fraction. We therefore retain 0.05 by the
locked selection rule, while treating the near-equal means as evidence that the
result is insensitive within this tested range rather than as a material gain
from the selected fraction.

We isolate this objective on a controlled \TrainingGaussians{}-million-Gaussian
sequence using three seeds, \TrainingSteps{} updates, and \TrainingWorlds{}
simultaneous worlds after warm-up. Reconstruction-only training reaches
13.98$\pm$0.22 dB on the clean state, 13.82$\pm$0.20 dB worst-10\% PSNR under
5\% IID loss, and 13.51$\pm$0.19 dB under burst loss. IID-only erasure training
does not close the tail gap: its corresponding values are 13.38$\pm$0.10,
13.39$\pm$0.09, and 13.27$\pm$0.03 dB. Burst-CVaR alone is worse at
12.96$\pm$0.11, 12.97$\pm$0.11, and 12.90$\pm$0.10 dB. The mixed objective with
24 clean anchors, CVaR, prefix monotonicity, and temporal flicker supervision is
the only variant that improves all three: 14.57$\pm$0.27 dB clean,
14.38$\pm$0.22 dB IID worst-10\%, and 13.91$\pm$0.27 dB burst worst-10\%. Its
allocator peak is 12.71 GiB versus 6.52 GiB for the single-world objectives,
which is the measured cost of retaining clean anchors and partial worlds in the
same H100 step. Under the severe bandwidth-plus-deadline condition every variant
collapses to 6.61 dB with a \TrainingDeadlineCliff{}\% cliff rate, so the result is
reported as partial-state robustness, not as a solution to missing base groups at
playout.

\section{Counterfactual Render Utility}
A refinement's usefulness is not constant: its image-space contribution depends on
which neighbouring clusters and prerequisite levels are already visible. For each
completion group $g$, we sample valid partial receiver contexts $m$ and render
paired states with and without $g$. Its marginal contribution is
\begin{equation}
\Delta_g(m)=D\!\left(I,R(m)\right)-D\!\left(I,R(m\cup g)\right),
\end{equation}
where $R$ is the differentiable renderer and $D$ combines perceptual and
pixel-space distortion. We retain the mean gain, lower-tail gain, and interaction
variance across contexts. A risk-adjusted utility
\begin{equation}
\widetilde\Delta_g=\frac{(1-\alpha)\,\mathbb{E}_m[\Delta_g]
+\alpha\,Q_{0.1}(\Delta_g)}{1+\beta\,\operatorname{Std}_m(\Delta_g)}
\end{equation}
favours groups that improve many incomplete render states rather than only the
complete scene. This quantity diagnoses brittle spatial and temporal regions and
supplies image-derived priorities to the delivery scheduler. The reported
frontiers use $\alpha=0.25$, $\beta=0.05$, and the 0.10 lower-tail quantile. A
flat utility table recovers the conventional scheduler, so this layer can be
ablated without changing the bitstream or decoder.

We call the resulting statement the \emph{Renderable-State Utility Principle}:
under incomplete delivery, the nominal order of a refinement is not an ordering of
its image-space utility. Utility must be conditioned on the renderable receiver
state. This is a testable claim about the representation, not a claim that every
fine refinement is harmful.

The held-out audit applies this diagnostic to D-NeRF \texttt{bouncingballs} and
HyperNeRF \texttt{broom2} and \texttt{chicken}, using four timestamps per scene.
The scene/context sizes are \RealUtilityClusters{} clusters with
\RealUtilityContexts{} contexts per group, \BroomUtilityClusters{} clusters with
\BroomUtilityContexts{} contexts per group, and \ChickenUtilityClusters{}
clusters with \ChickenUtilityContexts{} contexts per group, respectively. The
finest level has negative mean utility in \RealUtilityFineNegativeMean{},
\BroomUtilityFineNegativeMean{}, and \ChickenUtilityFineNegativeMean{} clusters,
while its lower-decile utility is negative in \RealUtilityFineNegativeTail{},
\BroomUtilityFineNegativeTail{}, and \ChickenUtilityFineNegativeTail{} clusters.
On the disjoint \texttt{broom2} training cameras, the corresponding counts are
\BroomTrainFineNegativeMean{} and \BroomTrainFineNegativeTail{} clusters. This
cross-view agreement supports the narrower principle: a group's nominal level
does not predict its contribution in a valid partial receiver context. Using a
normal approximation based on the context interaction standard error, the L3
mean-utility 95\% interval remains strictly below zero for
\RealUtilityFineCINegative{}, \BroomUtilityFineCINegative{}, and
\ChickenUtilityFineCINegative{} clusters, respectively. As a renderability check
rather than a codec comparison, the same \texttt{chicken} representation produces
four monotonic prefix points (\ChickenPrefixZeroPSNR{},
\ChickenPrefixOnePSNR{}, \ChickenPrefixTwoPSNR{}, and
\ChickenPrefixThreePSNR{}~dB) against a \ChickenSourcePSNR{}~dB source render,
with \ChickenPrefixViolations{} PSNR regressions.

\begin{figure}[H]
\centering
\includegraphics[width=0.98\linewidth]{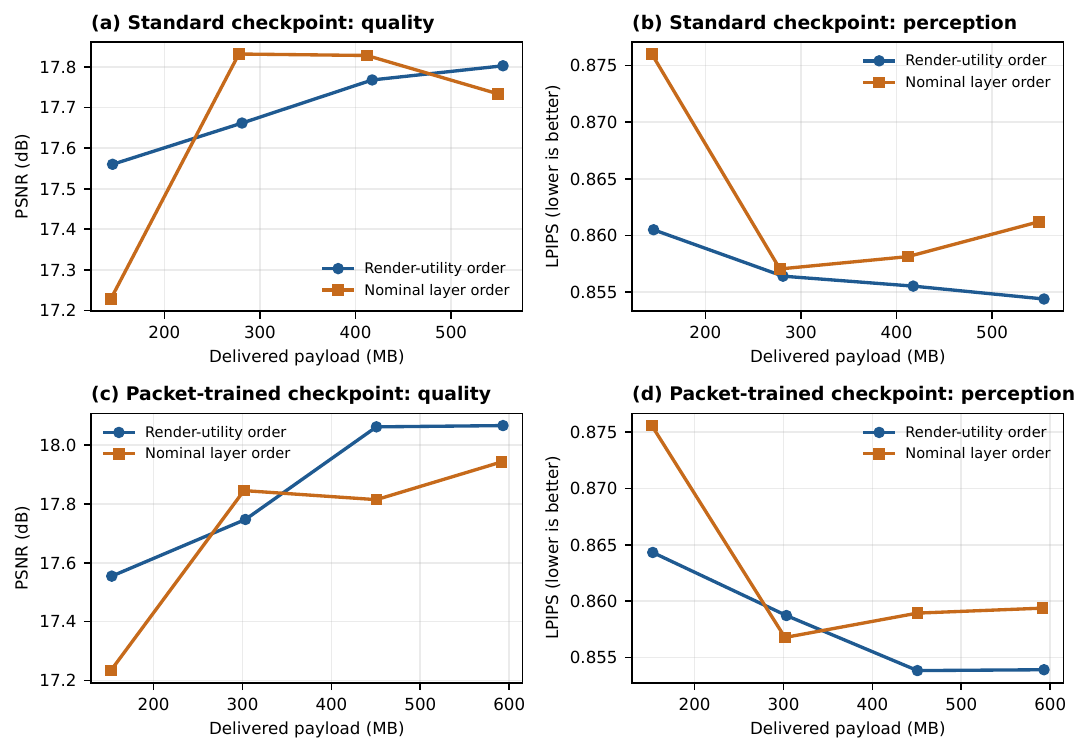}
\caption{Dependency-closed frontiers on held-out HyperNeRF \texttt{broom2}
cameras. Every nominal-layer point uses at least 98.7\% of the corresponding
utility-order payload budget. Utility order is monotonic in PSNR for both
checkpoints, whereas nominal layer order regresses despite receiving more bytes.}
\label{fig:frontier}
\end{figure}

Figure~\ref{fig:frontier} converts the measured utilities into a
dependency-closed scheduling frontier and evaluates every selected state on
held-out \texttt{broom2} cameras. At four matched byte budgets, utility ordering
has \FrontierUtilityViolations{} PSNR regressions versus
\FrontierLayerViolations{} for nominal layer order, whose payload utilization is
at least \FrontierLayerBudgetUse{}\%. At the lowest budget, utility ordering gains
\FrontierLowPSNRGain{} dB and reduces LPIPS by \FrontierLowLPIPSReduction{}\%; at
the final point the differences are \FrontierFinalPSNRGain{} dB and
\FrontierFinalLPIPSReduction{}\%. Repeating the test with the packet-trained
checkpoint changes the regression count from \RobustFrontierLayerViolations{} to
\RobustFrontierUtilityViolations{} and ends with
\RobustFrontierFinalPSNRGain{} dB higher PSNR and
\RobustFrontierFinalLPIPSReduction{}\% lower LPIPS. This is scheduling evidence,
not a claim of superior codec rate--distortion against external baselines.

\FloatBarrier

The controlled intervention in Figure~\ref{fig:frontier} changes only the
completion-group order; the checkpoint, dependency closure, renderer, and matched
payload budgets remain fixed. The observed regressions therefore arise from the
nominal refinement order rather than from decoding different bytes or evaluating
different scene models.

\begin{figure}[H]
\centering
\includegraphics[width=0.90\linewidth]{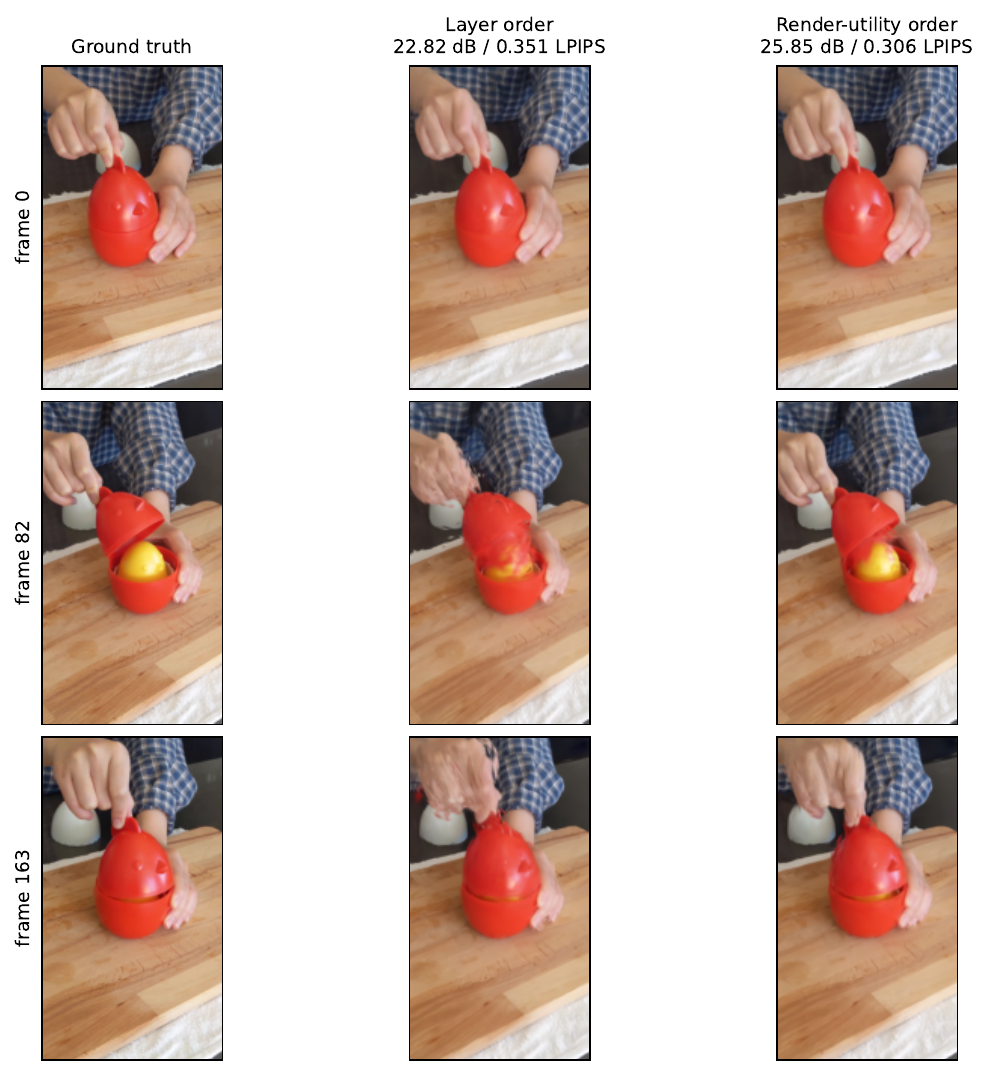}
\caption{Held-out HyperNeRF \texttt{chicken} frames at the lowest matched byte
budget. Utility is estimated on disjoint training cameras. Nominal layer order
spends \ChickenCrossviewLayerBudgetUse{}\% of the utility-order budget but
produces visible hand and object artifacts; render-utility order allocates the
same bytes to completion groups with higher marginal image contribution.}
\label{fig:crossview-utility}
\end{figure}

Figure~\ref{fig:crossview-utility} repeats the ordering intervention without view
leakage on HyperNeRF \texttt{chicken}. Utilities are measured only on training
cameras, while the selected states are rendered on held-out test cameras. At the
lowest matched budget, render-utility order gains
\ChickenCrossviewPSNRGain{} dB and reduces LPIPS by
\ChickenCrossviewLPIPSReduction{}\%, while nominal layer order consumes
\ChickenCrossviewLayerBudgetUse{}\% of the same budget. With the packet-trained
quantizer, the corresponding gains are \ChickenRobustCrossviewPSNRGain{} dB and
\ChickenRobustCrossviewLPIPSReduction{}\%. The frame strip tests whether these
aggregate differences correspond to localized hand and object artifacts. In the
rendered strip, the utility-ranked state preserves the manipulated object and
hand boundary more faithfully at the same byte budget. Because ranking and
evaluation use disjoint camera sets, this result does not depend on selecting
groups for the displayed views.

\FloatBarrier

\begin{figure}[H]
\centering
\includegraphics[width=0.99\linewidth]{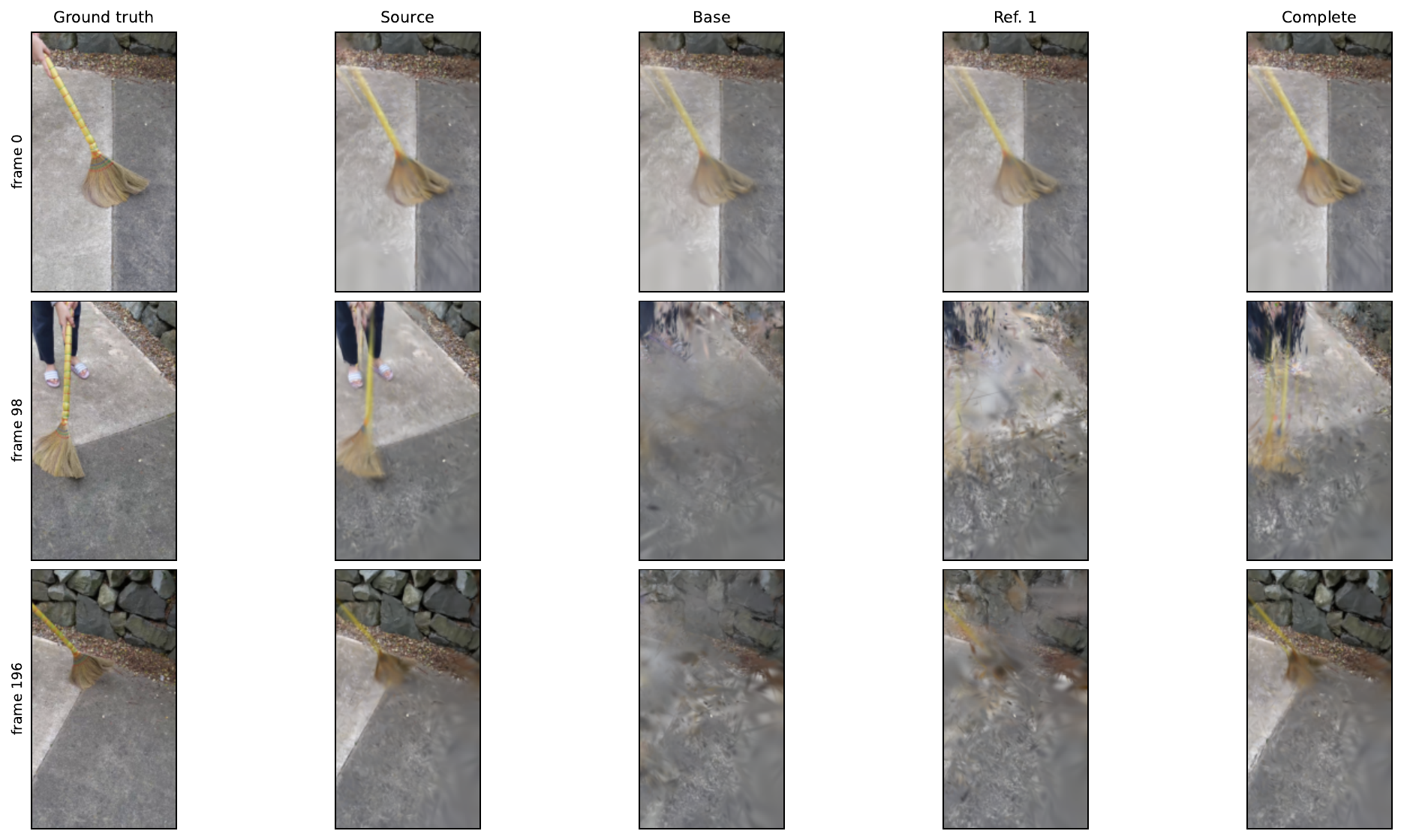}
\caption{Held-out HyperNeRF \texttt{broom2} frames at the complete decoded state.
This conservative failure-case diagnostic shows preserved coarse layout but weak
fine broom structure, disocclusion boundaries, and transient detail. It is not a
competitive visual-quality claim.}
\label{fig:realquality}
\end{figure}

The \texttt{broom2} comparison isolates the visual limits of the current
representation rather than the scheduling intervention. Figure~\ref{fig:realquality}
shows held-out views at the complete decoded state so that missing fine structure
cannot be attributed to packet loss or byte allocation. Coarse geometry and motion
remain recognizable, while thin broom fibers, disocclusion boundaries, and
transient appearance expose the present factorization error.

This diagnostic bounds the visual claim before the delivery experiments. The
complete decoded state preserves the coarse dynamic structure needed for a
renderable receiver representation, while the remaining artifacts identify where
the current factorization spends insufficient capacity. The erasure study below
therefore asks a narrower question: given this fidelity envelope, whether
packet-conditioned training changes the degradation slope when completion groups
are missing.

\FloatBarrier

\subsection{Erasure Robustness}

We next separate complete-state fidelity from delivery robustness by applying
identical completion-group erasures to the reference and packet-trained
quantizers. This paired protocol attributes any change in degradation slope to the
training objective rather than to different network samples.

\begin{figure}[H]
\centering
\includegraphics[width=0.98\linewidth]{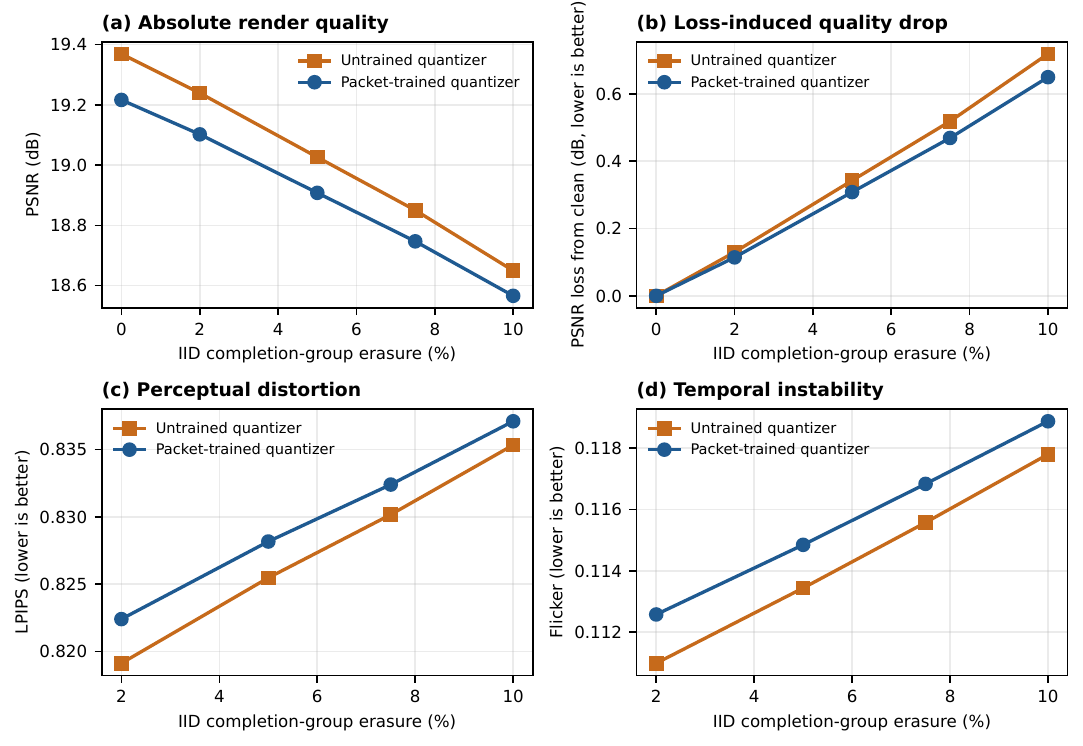}
\caption{Held-out HyperNeRF \texttt{broom2} response to IID completion-group
erasure. Absolute quality (a,c,d) exposes the clean-quality cost; clean-relative
PSNR loss (b) exposes the shallower degradation slope. Packet training is not
claimed to dominate until both views improve. Every panel includes the same
128-world seed and identical delivery masks for the two checkpoints.}
\label{fig:erasure-response}
\end{figure}

Figure~\ref{fig:erasure-response} makes the robustness cost explicit. Packet
training pays \ErasureCleanPSNRCost{} dB at the complete state. At 10\% IID
completion-group erasure, its loss relative to its own clean state is
\ErasureTenDropTrained{} dB rather than \ErasureTenDropUntrained{} dB, but its
absolute PSNR remains \ErasureTenAbsolutePSNRDeficit{} dB lower and LPIPS is
\ErasureTenLPIPSDeficit{}\% worse. Under the tested Gilbert--Elliott burst model,
the same checkpoint instead gains \ErasureBurstPSNRGain{} dB and improves LPIPS
by \ErasureBurstLPIPSGain{}\%. Thus the present training run flattens the
degradation slope and helps burst delivery, but does not yet dominate the
untrained quantizer under IID loss. A clean-heavier objective was also tested and
did not recover this gap; it is excluded from the selected method rather than
chosen post hoc. These results remain below the preregistered evidence threshold
and bound the current claim.

\FloatBarrier

The paired \texttt{chicken} test checks whether the \texttt{broom2} trend transfers
to a second scene. A 1,200-step run with 128 candidate worlds and 80 rendered
worlds per update improves the complete prefix by \ChickenTrainedCleanGain{} dB
and retains \ChickenTrainedPrefixViolations{} monotonicity violations. Under the
same 128 IID-5\% delivery masks, it wins all \ChickenPairedWinPercent{}\% of
paired worlds and gains \ChickenPairedPSNRGain{} dB (paired bootstrap 95\% CI
[\ChickenPairedPSNRCILow{}, \ChickenPairedPSNRCIHigh{}]). LPIPS, temporal LPIPS,
and flicker improve by \ChickenPairedLPIPSGain{}\%,
\ChickenPairedTemporalLPIPSGain{}\%, and \ChickenPairedFlickerGain{}\%,
respectively. The direction is consistent but the LPIPS magnitude remains far
below the 15\% robustness threshold; packet-conditioned quantization alone is
therefore insufficient for the broad robustness claim. Under 128 matched
Gilbert--Elliott burst worlds, the same checkpoint gains
\ChickenBurstPSNRGain{} dB (95\% CI [\ChickenBurstPSNRCILow{},
\ChickenBurstPSNRCIHigh{}]) and improves LPIPS, temporal LPIPS, and flicker by
\ChickenBurstLPIPSGain{}\%, \ChickenBurstTemporalLPIPSGain{}\%, and
\ChickenBurstFlickerGain{}\%. However, the fraction of worlds more than 3 dB
below the clean state changes from \ChickenBurstReferenceCliff{}\% to
\ChickenBurstConditionedCliff{}\%. We therefore report the paired mean gain but
do not claim removal of cliff failures.

\begin{figure}[H]
\centering
\includegraphics[width=\linewidth]{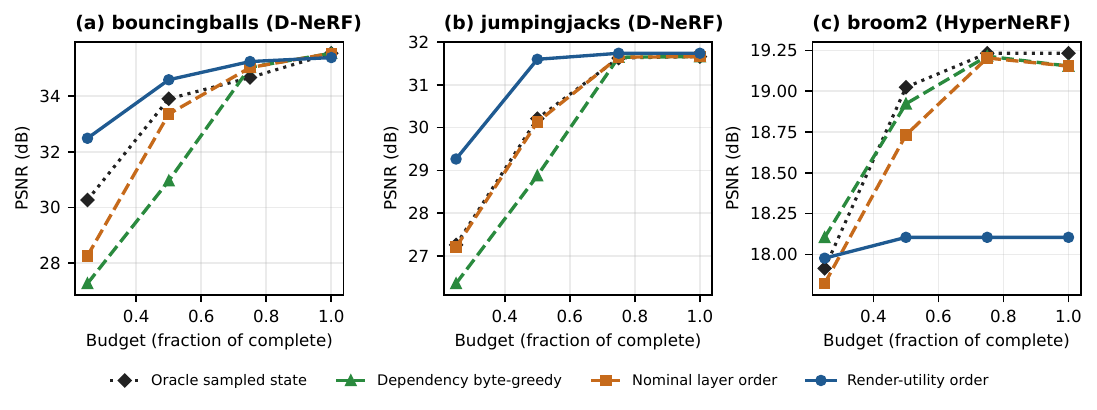}
\caption{Partial-state oracle atlas. For each scene, PSNR of three scheduling
policies and the best sampled dependency-closed state (oracle) across four matched
byte budgets, over \AtlasStatesEvaluated{} sampled states per scene. Render-utility
order matches or exceeds the sampled oracle on the D-NeRF scenes but plateaus below
it on \texttt{broom2}, where the utility signal is less reliable.}
\label{fig:atlas}
\end{figure}

The frontier in Figure~\ref{fig:frontier} fixes a single delivery order. The
partial-state oracle atlas instead samples \AtlasStatesEvaluated{}
dependency-closed receiver states per scene and records, at each of
\AtlasBudgetCount{} matched byte budgets, the best sampled state as an empirical
oracle alongside three scheduling policies: render-utility order, nominal layer
order, and dependency byte-greedy order. Figure~\ref{fig:atlas} reports per-scene
PSNR. On the two D-NeRF scenes the render-utility order is the strongest policy
and matches or exceeds the sampled oracle at the tightest budget
(\texttt{bouncingballs} \AtlasBBUtilityLow{}~dB and \texttt{jumpingjacks}
\AtlasJJUtilityLow{}~dB, against sampled oracle \AtlasBBOracleLow{} and
\AtlasJJOracleLow{}~dB and nominal layer order \AtlasBBNominalLow{} and
\AtlasJJNominalLow{}~dB). Because the oracle is the best of thousands of
\emph{sampled} states, a constructive policy that exploits the utility signal can
surpass it. On the harder HyperNeRF \texttt{broom2} scene the same policy is
monotone but conservative: it plateaus at \AtlasBroomUtilityTop{}~dB,
\AtlasBroomOracleGapTop{}~dB below the oracle, while nominal layer and byte-greedy
orders track the oracle. The render-conditioned utility navigates the partial-state
space well on the synthetic dynamic scenes, but its reliability degrades on this
real captured scene, consistent with the scoped claims of this work.

\FloatBarrier

\section{GPU-Batched Counterfactual Rendering}
The same learned scene is evaluated under many delivery masks. \texttt{gsplat}
supports leading batch dimensions for Gaussians and cameras~\citep{gsplat2025}, so
states with shape $[W,N,\cdot]$ are rasterized in one call. Learned components run
in BF16 where stable, while rasterization-sensitive operations remain FP32. A
probe-based tuner finds the largest world batch under a target memory budget and
recovers from OOM probes.

Our reporting follows established benchmark practice: throughput is separated
from latency behavior~\citep{mlperfInference2020}, stochastic measurements are
repeated rather than represented by one selected run~\citep{mlperfTraining2020},
and the complete iteration distribution is retained. We report three independent
repetitions per world batch, Student-$t$ 95\% confidence intervals for throughput,
and empirical CDFs for forward-plus-backward iteration time. Device utilization,
resident memory, power, and energy are sampled independently of PyTorch allocator
accounting. Each timed region discards five warmup iterations, and every
measurement uses the software stack pinned in \texttt{environment.lock.json}: an
NVIDIA H100 PCIe (compute capability~9.0, driver~570.158.01), CUDA~12.8,
PyTorch~2.11.0, \texttt{gsplat}~1.5.3, and Python~3.12.13. The benchmark driver,
world-batch tensor layout $[W,N,\cdot]$, delivery masks, and random seed are fixed
across repetitions, so the reported intervals reproduce from the released harness.

\begin{figure}[H]
\centering
\includegraphics[width=0.98\linewidth]{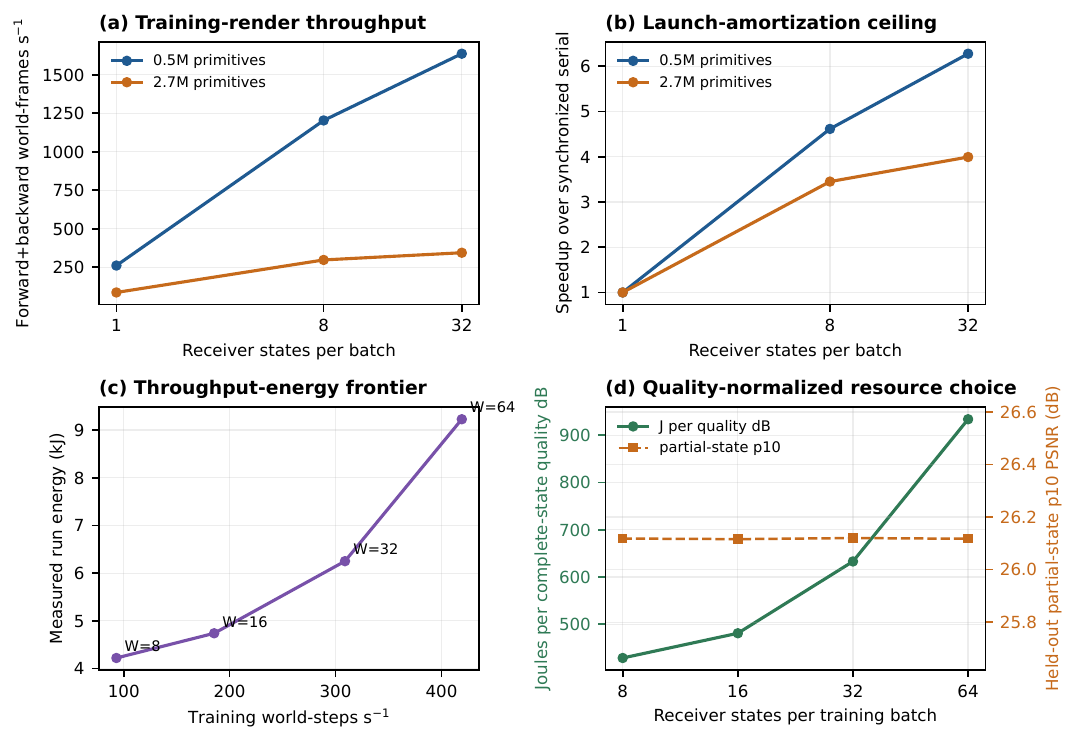}
\caption{H100 PCIe benchmark evidence. Panels show training-render throughput,
synchronized-serial speedup, the throughput-energy frontier, and the
quality-normalized resource choice. The p10 partial-state PSNR line in panel (d)
is plotted on an honest one-dB axis to show that tail quality is essentially
invariant across the measured world-batch choices.}
\label{fig:h100}
\end{figure}

The forward-plus-backward roofline sweep separates training throughput from the
small forward-only stress test. With 2.7M primitives, 32 receiver states provide
\RooflineLargeSpeedupThirtyTwo{}$\times$ speedup over synchronized serial
evaluation and reach only \RooflineLargeBandwidthPercent{}\% of the measured
streaming-bandwidth roof under the conservative traffic model. The bottleneck is
therefore launch and rasterization overhead, not H100 memory bandwidth. A separate
resident-world stress test holds \ResidentGaussians{} million Gaussians across
each of 128 worlds, reaches \ResidentVRAM{} GB peak allocation, and records
\ResidentUtilMean{}\% mean and \ResidentUtilPctl{}\% 95th-percentile active GPU
utilization. Mean active power is \ResidentPower{} W and the measured run consumes
\ResidentEnergyKJ{} kJ on an H100 PCIe.

The utility experiment evaluates \UtilityStates{} rendered world states in
\UtilitySeconds{} s (\UtilityThroughput{} states/s) with \UtilityVRAM{} GiB peak
allocator residency. Unlike the synthetic allocator stress test, these worlds are
paired image-space interventions used directly by the scheduler study.

\FloatBarrier
\section{Network Delivery Realization}
The representation does not require a particular transport. Our implementation
uses reliable control metadata and deadline-sensitive media datagrams under one
congestion controller and pacer. The receiver exposes the renderable dependency
closure rather than raw packet arrivals.

\subsection{Channels and feedback}
The sender exposes three logical channels. First, a reliable control stream carries
the DGSM manifest and codec version. Second, media datagrams carry DGSP chunks.
Third, feedback reports acknowledgement ranges, recovered packet identifiers,
completed cluster-layer groups, and receiver playout time. Reliable and unreliable
data consume the same congestion budget.

\subsection{Dependency-completion scheduling}
For a cluster-layer group $g$, let $C(g)$ be all unacknowledged chunks in $g$ plus
the transitive dependency closure needed to render it. The scheduler estimates
deadline delivery probability $P_g$ from remaining slack, path rate, RTT, and loss,
then ranks feasible groups by
\begin{equation}
S(g)=\frac{P_g\,\widetilde\Delta_g\,U_g}{\sum_{p\in C(g)}B_p},
\end{equation}
where $B_p$ is wire size and $U_g$ increases as an incomplete group approaches
completion. Selecting completion groups prevents spending bytes on partial groups
that cannot change the rendered state. A layered scheduler, a segment-budget
DASH-style scheduler, and a dependency-aware scheduler without deadline utility
are controlled baselines.

\subsection{Congestion control and pacing}
The emulator implements RFC-9002-style NewReno accounting: slow start, additive
increase, multiplicative decrease, bytes in flight, smoothed RTT, RTT variance, and
a probe timeout
\begin{equation}
\mathrm{PTO}=\mathrm{SRTT}+\max(4\,\mathrm{RTTVAR},1\mathrm{ms})+d_{ack}.
\end{equation}
The application pacer is capped by both congestion state and estimated path
capacity, $r_{send}=\min(r_{cwnd},\widehat r_{path})$. Packet loss timers begin
after serialization, not at application enqueue time. Reordering and finite queues
are explicit trace parameters.

The robustness sweep also includes a BBR-inspired model controller. It estimates
delivery rate from acknowledged bytes, tracks the minimum RTT over the replay
window, targets a two-BDP inflight budget, and applies a loss-derived inflight
ceiling after burst events. This is an ablation model for scheduler sensitivity,
not a production BBRv2 implementation.

\subsection{Selective recovery and receiver closure}
The compact manifest is retransmitted reliably. Base chunks receive a small
retransmission budget because losing them invalidates refinements. Refinements are
not retransmitted after their usefulness deadline. Optional systematic XOR repair
protects configurable base groups and can recover one missing datagram; the result
must pass the original DGSP CRC.

The receiver buffers reordering, validates CRCs, performs repair, and records raw
arrivals separately from usable packets. A cluster-layer is usable only when all
chunks in its dependency closure are usable. At each playout timestamp it snapshots
this closure and renders without waiting for expired data. Bytes outside the usable
closure are counted as wasted goodput.

\subsection{Trace-driven delivery results}
\begin{figure}[H]
\centering
\includegraphics[width=0.90\linewidth]{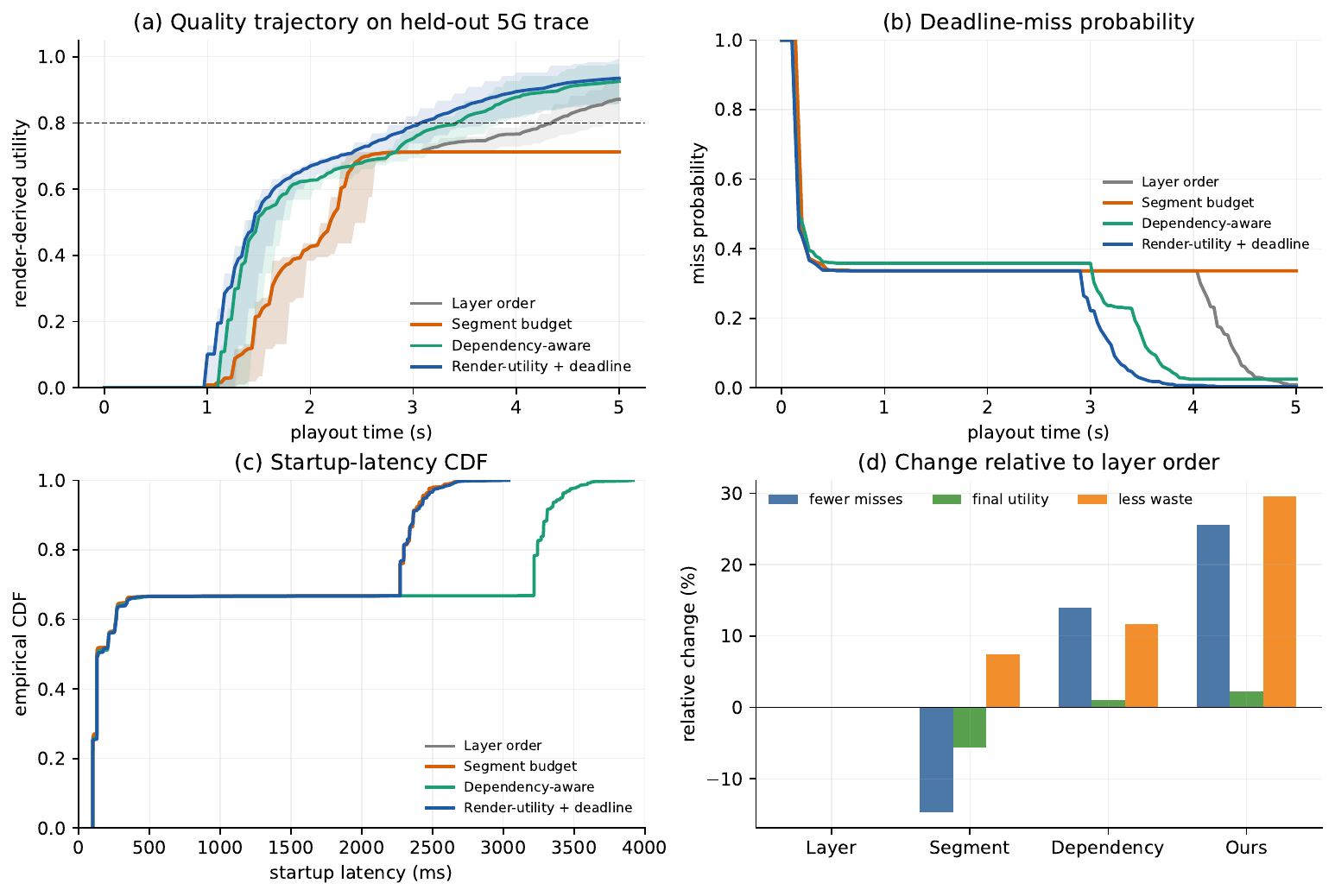}
\caption{Trace-driven QoE under identical packet graphs and path realizations.
Bands in (a) are 95\% empirical intervals over 200 seeds. Panels report quality
over playout time, deadline-miss probability, startup CDF, and relative changes on
measured mobile traces. Utility is derived from paired renders, not a layer index.}
\label{fig:transportqoe}
\end{figure}

We replay the same packet graph under layer order, segment-budget adaptation,
dependency-aware scheduling, and the proposed render-utility-plus-deadline policy.
The complete sweep contains \TransportSimulations{} simulations over
\TransportTraceCount{} held-out synthetic and measured traces and retains
\TransportTrajectoryRows{} frame-level QoE records. Figure~\ref{fig:transportqoe}
shows the layered-dependency, base-FEC, burst-loss condition; every policy shares
the same congestion controller, packet losses, queue, and random seed.

\begin{table}[H]
\centering
\caption{Transport-policy ablation on measured mobile traces with layered
dependencies, burst loss, and four-packet base repair groups. Lower miss and
waste are better; utility is render-derived.}
\label{tab:transportpolicy}
\begin{tabular}{llrrrr}
\toprule
Controller & Scheduler & Startup (ms) & Miss (\%) & Utility & Waste (MiB) \\
\midrule
bbr-model & dash & 910.5 & 51.5 & 0.854 & 0.004 \\
bbr-model & deadline-utility & 912.8 & 24.6 & 0.990 & 0.002 \\
bbr-model & dependency & 1471.5 & 21.9 & 0.993 & 0.002 \\
bbr-model & layered & 910.5 & 31.7 & 0.993 & 0.004 \\
newreno & dash & 931.0 & 51.6 & 0.849 & 0.004 \\
newreno & deadline-utility & 936.0 & 24.0 & 0.985 & 0.002 \\
newreno & dependency & 1500.5 & 21.8 & 0.988 & 0.002 \\
newreno & layered & 931.0 & 31.8 & 0.988 & 0.004 \\
\bottomrule
\end{tabular}

\end{table}

\FloatBarrier

Across \TransportRealPairs{} paired seed--trace trials on UCC 5G and two
CellReplay-derived traces, render-utility scheduling reduces deadline misses by
\TransportMissReduction{}\% (bootstrap 95\% CI
\TransportMissReductionLow{}--\TransportMissReductionHigh{}\%) relative to layer
order. Final render-derived utility rises by \TransportQualityGain{}\% and wasted
goodput falls by \TransportWasteReduction{}\%. This is not a free startup win:
mean finite startup increases by \TransportStartupCost{} ms, and
\TransportStartupFailures{} of the 600 proposed-policy trials fails to reach the
startup threshold within the replay window. The trajectory and miss curves make
that tradeoff visible instead of hiding it in a terminal average.

Table~\ref{tab:transportpolicy} separates the scheduling decision from the
congestion controller. With NewReno, deadline-utility scheduling cuts misses from
31.8\% to 24.0\% relative to layered order and halves wasted goodput from 0.004
to 0.002 MiB. With the BBR-inspired controller the same comparison is 31.7\% to
24.6\% miss with the same waste reduction. Dependency-only scheduling attains the
lowest miss rate, 21.8--21.9\%, but delays startup to 1.47--1.50 s because it
waits for more closure before playback. Figure~\ref{fig:transportqoe} therefore
plots the time trajectory rather than only the terminal utility score.

\FloatBarrier

The trace result establishes the scheduling side of the formulation under a
fixed packet graph. Image-derived utility changes which delivered bytes become
renderable before their display deadlines, but it does not by itself establish a
better scene representation. The evaluation protocol therefore separates three
evidence layers: graphics quality on real scenes, oracle partial-state geometry,
and trace-driven delivery behavior.

\FloatBarrier

\section{Evaluation Protocol}
\subsection{Graphics evaluation}
The real-scene study is locked before comparative training. It uses D-NeRF
\texttt{bouncingballs} and \texttt{jumpingjacks}, HyperNeRF \texttt{broom2} and
\texttt{chicken}, and frames 0--29 of 8i \texttt{longdress} and \texttt{soldier}.
Baselines include the uncompressed deformation-field implementation of Wu et al.,
Dynamic-LapisGS, uniform quantization plus
rANS, context-adaptive rANS, Light4GS, QUEEN, an LTS-style DASH scheduler,
low-delay HEVC/AV1, and executable progressive-decomposition and GS-NFS releases or explicitly labelled
reimplementations.

The primary endpoints are rate--distortion, LPIPS, temporal LPIPS, flicker, and
monotonicity over valid partial states. Every scene includes matched-camera frame
strips, perceptual-error heatmaps, temporal-difference maps, disocclusion crops,
and synchronized videos. Nonrigid motion, topology change, and transient appearance
are explicit failure categories. Independent versus layered chunks, affine versus
low-rank motion, refinement count, context entropy coding, erasure training, CVaR,
and counterfactual utility are isolated at equal wire rate.

\subsection{Partial-state oracle atlas}
To separate representation quality from any particular scheduler, we also build a
partial-state oracle atlas for each completed real scene. The atlas samples
thousands of dependency-closed receiver states, evaluates them on held-out
cameras, and selects the best sampled state under each byte budget by PSNR and by
LPIPS. Nominal layer order, dependency-closed byte-greedy order, and
render-utility order are then compared to this sampled oracle with identical
decoder state and renderer. The same pass records the worst lower-tail states as
frame strips and error maps. This experiment uses the H100 renderer to map the
image-space geometry of incomplete dynamic Gaussian states rather than only to
measure throughput.

\subsection{Network evaluation}
Transport evaluation uses identical packet graphs and traces across policies. It
includes synthetic 2--200 Mbit/s paths, 10--320 ms RTT, 0--10\% IID loss,
Gilbert--Elliott bursts, UCC 5G traces~\citep{raca2020fiveg}, and CellReplay
scenarios~\citep{cellreplay2025}. UCC bitrate, ping, jitter, and loss are replayed
directly. CellReplay-derived scalar traces count native 1400-byte heavy-PDO
delivery opportunities in 100 ms bins and use measured base delay; native
\texttt{mm-cellular} replay remains a separate end-to-end experiment.

We report quality as a function of wall-clock playout time, not only aggregate
PSNR. Metrics include startup-to-quality trajectory, PSNR, SSIM, LPIPS, temporal
flicker, quality switches, deadline-miss and stall curves, burst-recovery time,
received and usable goodput, decode/render throughput, peak memory, and energy.
Three seeds and bootstrap 95\% confidence intervals are required.

Network ablations independently remove completion-group scheduling, dependency
closure, congestion-window interaction, burst loss, retransmission, FEC, and
deadline feasibility. Deadline thresholds and adaptation intervals are swept under
identical traces. This tests whether graphics-side robustness survives a concrete
delivery stack; it is not a substitute for representation quality.

\FloatBarrier
\section{Computational Validation}
The H100 sweep contains eight world-batch settings and three independent
repetitions per setting. At 128 worlds it reaches \SweepThroughput{}
world-frames/s, \SweepUtilization{}\% mean active utilization,
\SweepMemoryGiB{} GiB mean device residency, and \SweepPower{} W mean active
power. Throughput begins to saturate after 32 worlds while memory continues to
grow, so the largest resident batch is not automatically the most efficient
training point. This distinction is important when selecting the final scene-wise
training configuration.

\begin{table}[H]
\centering
\caption{H100 world-batch benchmark. Throughput confidence intervals use the
three independent repetitions for each world count. A world-frame is one rendered
receiver-state image for one scene timestamp.}
\label{tab:h100benchmark}
\begin{tabular}{rrrrrr}
\toprule
Worlds & Throughput & Residency & Util. & Power & Energy \\
 & (wf/s, 95\% CI) & (GiB) & (\%) & (W) & (J/iter) \\
\midrule
1 & 72.4$\pm$1.3 & 1.7 & 93.7 & 91.3 & 2.2 \\
2 & 129.4$\pm$0.1 & 2.2 & 96.0 & 107.4 & 2.5 \\
4 & 201.1$\pm$0.2 & 3.3 & 94.7 & 116.8 & 2.5 \\
8 & 271.1$\pm$0.9 & 5.4 & 96.3 & 160.5 & 5.0 \\
16 & 310.9$\pm$0.5 & 9.5 & 98.0 & 207.9 & 9.3 \\
32 & 333.9$\pm$0.7 & 17.8 & 98.6 & 237.9 & 21.7 \\
64 & 352.2$\pm$0.1 & 34.3 & 94.9 & 257.4 & 50.2 \\
128 & 360.5$\pm$0.3 & 67.3 & 98.2 & 279.5 & 100.4 \\
\bottomrule
\end{tabular}

\end{table}

The test suite covers rANS and context-rANS round trips, packet and manifest CRCs,
MTU enforcement, dependency loss, partial decode, XOR repair, congestion-window
response, low-capacity pacing, trace lookup, QoE emission, and CUDA batched/serial
equivalence. Codec accounting separately reports raw symbols, entropy lower bound,
rANS gap, chunking loss, fixed header, dependency identifiers, CRC, compact
manifest, cluster metadata, and total wire bits per Gaussian. These quantities make
the price of independent packet renderability explicit.

\begin{table}[H]
\centering
\caption{Per-scene codec accounting for D-NeRF \texttt{bouncingballs}. The table
separates entropy cost, MTU chunking, dependency identifiers, CRC, and compact
control metadata so the cost of independent packet renderability is visible.}
\label{tab:codecaccounting}
\begin{tabular}{lrr}
\toprule
Component & MiB & Role \\
\midrule
Raw quantized symbols & 68.839 & -- \\
Empirical entropy bound & 42.817 & -- \\
Monolithic rANS & 42.821 & reference \\
MTU-chunked rANS payload & 42.942 & payload \\
Fixed packet headers & 0.792 & overhead \\
Dependency identifiers & 0.141 & overhead \\
CRC & 0.144 & overhead \\
Compact control manifest & 0.314 & startup \\
Total wire plus manifest & 44.334 & total \\
\midrule
Packetization overhead & 2.45\% & wire \\
Robustness premium & 3.53\% & vs. monolithic \\
Wire bits/Gaussian & 12894.7 & total \\
\bottomrule
\end{tabular}

\end{table}

The accounting table makes the byte cost of independent renderability explicit.
Its role is diagnostic: if later scene-level RD curves fail, the table identifies
whether the loss comes from entropy coding, MTU chunking, dependency metadata, or
the representation itself. This closes the computational validation by tying the
systems measurements to the codec cost model used in the gates below.

\section{Limitations and Completion Criteria}
The principal scientific result remains incomplete until the six real scenes and
strongest runnable baselines are measured. Fixed Gaussian identity may fail under
topology change, the affine plus compact motion basis may underfit nonrigid motion,
and XOR repair is intentionally simpler than production fountain or Raptor codes.
The transport emulator follows QUIC-facing semantics but does not replace a native
QUIC implementation or CellReplay's workload-sensitive emulator.

Broad comparative superiority is treated as a locked evidence threshold, not as a
result of the current experiments. The threshold requires at least 15\% average
BD-rate reduction under Bjontegaard integration~\citep{bjontegaard2001} against
the strongest runnable dynamic Gaussian baseline, at least 15\% LPIPS reduction
at 5\% loss~\citep{zhang2018perceptual}, monotone rate--quality points from one
model, and removal of cliff-like quality failures. Graphics evidence must include
frame strips, error heatmaps, temporal-difference visualizations, side-by-side
videos, and failure cases. Until these criteria are met, the claim is restricted
to render-conditioned utility and delivery-state robustness diagnostics.

\section{Reproducibility}
Original code and documentation are licensed under CC BY 4.0. Third-party
repositories and datasets are not vendored. Pinned commits, checksums, converter
assumptions, locked configs, packet and frame schemas, H100 measurements, and
artifact checks are included. Dataset archives are processed one scene at a time
and removed after durable synchronization to respect the worker disk budget. The
arXiv source archive is self-contained for TeX compilation and artifact inspection.
Its root contains \texttt{main.tex}, the bibliography, generated numeric macros,
the referenced PDF figures, generated table sources, and the CC BY 4.0 license.
The \texttt{anc/} directory contains the implementation source, tests, experiment
and packaging scripts, immutable configurations, dependency pins, third-party
patches, environment metadata, packet and frame schemas, and the JSON/CSV records
from the reported measurements. The archive contains no credentials, private
keys, model checkpoints, raw dataset files, dataset-derived point clouds, vendored
third-party repositories, TeX build products, interpreter caches, test caches, or
runtime logs.

\begingroup
\small
\bibliographystyle{plainnat}
\bibliography{references}
\endgroup

\section*{Appendix: Worker Storage and Source-Archive Protocol}
The H100 worker used for the real-scene runs exposes fast GPU memory and system
RAM but only a small persistent filesystem. The evaluation therefore treats disk
space as a scheduled resource. Raw scene archives, exported PLY sequences, and
large transient checkpoints are never allowed to accumulate across scenes on the
overlay filesystem. Inputs are staged to RAM with a checksum manifest, the run
consumes the RAM path, compact JSON/CSV/PDF/MP4 evidence is synchronized to
durable storage, and raw scene material is removed before the next scene starts.
Large transient checkpoints that must remain addressable during a worker session
are spilled to RAM and replaced by symlinks at their original paths; this preserves
script compatibility while freeing persistent space.

The worker cycle is implemented as Algorithm~\ref{alg:worker}.
\begin{algorithm}[H]
\renewcommand{\thealgorithm}{A1}
\caption{RAM-staged scene cycle for a low-disk H100 worker}
\label{alg:worker}
\begin{algorithmic}[1]
\Require locked scene order $S$; persistent root $P$; RAM root $M$; durable
evidence root $D$; safety floor $F$
\For{scene $s \in S$}
  \State fetch or unpack only $s$ into $P/s$; write \textsc{sha256} manifest $h_s$
  \State copy $P/s$ to $M/s$; verify \textsc{sha256}$(M/s)=h_s$
  \If{free space on $P < F$}
    \State \textbf{abort} before training
  \EndIf
  \State run training and evaluation with input root set to $M/s$
  \If{a transient checkpoint must remain addressable}
    \State move it to $M/\mathrm{spill}$ and leave a symlink at the original path
  \EndIf
  \State synchronize JSON, CSV, \TeX{} table inputs, PDF/PNG figures, and
  compressed videos to $D/s$
  \State remove raw archives, exported PLY sequences, dataset images, and $M/s$
  \State assert free space on $P \ge F$ before advancing to the next scene
\EndFor
\end{algorithmic}
\end{algorithm}
The scene-cycle script implements this pattern. The data-path verifier checks
that the run observes the RAM path, checksum verification is present, durable
outputs survive, and both disk and RAM inputs are cleaned afterward. This is why
the data-path gate is a machine-checked artifact rather than a prose assumption.

The source archive follows the
\href{https://info.arxiv.org/help/ancillary_files.html}{arXiv ancillary-file
placement rule}. The zip root contains the TeX source, bibliography, generated
numeric macros, required figures, tables, and license. The anc directory contains
code, tests, locked configs, compact measurement records, schemas, and upstream
patch metadata. No TeX or BibTeX source is placed in anc; no script depends on
being run from an anc path; and the archive verifier rejects credentials, private
keys, model checkpoints, raw datasets, dataset-derived point clouds, vendored
external repositories, TeX build products, Python bytecode, caches, runtime logs,
README files, and PDF JavaScript or automatic actions.
\end{document}